\documentclass[conference,letterpaper]{IEEEtran}
\usepackage[letterpaper, left=0.75in, right=0.75in, bottom=0.75in, top=1in]{geometry}
\usepackage{cite}  
\usepackage[pdftex]{graphicx}
\graphicspath{{../pdf/}{../jpeg/}}
\DeclareGraphicsExtensions{.pdf,.jpeg,.png}

\makeatletter

\def\ps@IEEEtitlepagestyle{%
  \def\@oddfoot{\mycopyrightnotice}%
  \def\@evenfoot{}%
}
\def\mycopyrightnotice{%
  {\footnotesize To appear in Proc. ISBI 2021, April 13-16, 2021, Nice, France ~\hfill}% <--- Change here
  \gdef\mycopyrightnotice{}
}

\usepackage[cmex10]{amsmath}
\usepackage{mathabx}
\usepackage{algorithmic}
\usepackage{array}
\usepackage{mdwmath}
\usepackage{mdwtab}
\usepackage{eqparbox}
\usepackage{url}
\usepackage{lineno,hyperref}
\usepackage{booktabs}
\usepackage{multirow}
\let\labelindent\relax
\usepackage{enumitem}
\usepackage{amsmath,bm}
\usepackage{xcolor}
\modulolinenumbers[5]

\newcommand{\beq}{\begin{equation}}
	\newcommand{\eeq}{\end{equation}}

\begin{document}
	
	\title{\LARGE Modeling of Textures to Predict Immune Cell Status and Survival of Brain Tumour Patients}
	
	% \author{\authorblockN{Leave Author List blank for your IMS2013 Summary (initial) submission.\\ IMS2013 will be rigorously enforcing the new double-blind reviewing requirements.}
	% \authorblockA{\authorrefmark{1}Leave Affiliation List blank for your Summary (initial) submission}}
	\author{\IEEEauthorblockN{Ahmad Chaddad\IEEEauthorrefmark{1}, Mingli Zhang\IEEEauthorrefmark{2}, Lama Hassan\IEEEauthorrefmark{1} and Tamim Niazi\IEEEauthorrefmark{3}}
		\IEEEauthorblockA{
			\IEEEauthorrefmark{1} 
			School of Artificial Intelligence, Guilin University of Electronic Technology
			Guilin, Guangxi, China\\ 
			\IEEEauthorrefmark{2} Montreal Neurological Institute, McGill University, Montreal, Canada\\
			\IEEEauthorrefmark{3} Lady Davis Institute for Medical Research, McGill University, Montreal, Canada\\
			Email: ahmadchaddad@guet.edu.cn}}

	\maketitle
	
	\begin{abstract}
		Radiomics has shown a capability for different types of cancers such as glioma to predict the clinical outcome. It can have a non-invasive means of evaluating the immunotherapy response prior to treatment. However, the use of deep convolutional neural networks (CNNs)-based radiomics requires large training image sets. To avoid this problem, we investigate a new imaging features that model distribution with a Gaussian mixture model (GMM) of learned 3D CNN features. Using these deep radiomic features (DRFs), we aim to predict the immune marker status (low versus high) and overall survival for glioma patients. We extract the DRFs by aggregating the activation maps of a pre-trained 3D-CNN within labeled tumor regions of MRI scans that corresponded immune markers of 151 patients. Our experiments are performed to assess the relationship between the proposed DRFs, three immune cell markers (Macrophage M1, Neutrophils and T Cells Follicular Helper), and measure their association with overall survival. Using the random forest (RF) model, DRFs was able to predict the immune marker status with area under the ROC curve (AUC) of 78.67, 83.93 and 75.67\% for Macrophage M1, Neutrophils and T Cells Follicular Helper, respectively. Combined the immune markers with DRFs and clinical variables, Kaplan-Meier estimator and Log-rank test achieved the most significant difference between predicted groups of patients (short-term versus long-term survival) with p\,=\,4.31$\times$10$^{-7}$ compared to p\,=\,0.03 for Immune cell markers, p\,=\,0.07 for clinical variables , and p\,=\,1.45$\times$10$^{-5}$ for DRFs. Our findings indicate that the proposed features (DRFs) used in RF models may significantly consider prognosticating patients with brain tumour prior to surgery through regularly acquired imaging data.
	\end{abstract}
	
	\IEEEoverridecommandlockouts
	
	\IEEEpeerreviewmaketitle

	% ===================
	% # I. Introduction #
	% ===================
	
	\section{Introduction}
	Gliomas, the most common brain-initiating tumours, are classified by histopathological process into four grades (I, II, III or IV), depending on their aggressiveness, as noted by the World Health Organization (WHO) \cite{stupp2006changing}. Specifically, grade I is a non-invasive tumours and considered as generally benign, grade II and III are known as lower grade glioma (LGG) while grade IV represents the high grade glioma that is the most malignant tumours such as glioblastoma multiforme (GBM). GBM is the most deadly brain tumor with a median survival of 15 months \cite{sizoo2014measuring} while LGG patients is longer survival with an average approximately 7 years. At the time of diagnosis, most patients initially receive a surgical resection / biopsy and then radiation therapy (XRT) and/or the single chemotherapy agent temozolamide (TMZ) \cite{daniel2019temozolomide}. Many of these treated patients are leading to recurrence and death faster than others. Immmnotherapy is a remarkable cancer treatment approach that uses the patient’s own immune system to kill cancer cells \cite{brown2018harnessing}. This strategy is indicated that  the intratumoral immune response is related to the tumor progression and prognosis in gliomas \cite{yang2010cd8+}. Notwithstanding the development in treatment improvements, foreseeing reaction to immunotherapy preceding treatment stays a difficult errand for clinicians.
	
	With this favorable position of treatment, the identification of the response to immune therapeutics requires an invasive technique by either biopsy or medical procedure strategy, considering the natural dangers, restrictions of testing, difficulties and expenses related with biopsies. Therefore, a non-invasive biomarker is required to avoid these limitations. Radiomics is a non-invasive approach that is able to convert the images into quantitative data \cite{chaddad2019radiomics}. It examined the relationships between tumor imaging features, genomics, proteomics, transcriptomic, and clinical outcome \cite{zanfardino2019bringing,chaddad2019integration,zhou2019machine,chaddad2019deep}. Moreover, radiomics signatures has been used for assessing CD8 cell count and its corresponding relation to clinical outcomes of patients treated with immunotherapy \cite{sun2018radiomics}. For example, the tumour volume ratio in T2-FLAIR scans relative to the contrast enhancement volume was shown to be related to GBM mesenchymal subtype \cite{naeini2013identifying} that has a stronger immunological response compared to other GBM subtypes \cite{doucette2013immune}. Recent work has examined the relationships between tumor imaging features (e.g., shape, texture, histogram), immune cell markers and the corresponding survival outcome of patients with glioma \cite{chaddad_deep_nodate,chaddad2019deep2}.
	
	To date, limited studies have been considered the deep CNNs in the context of radiomics. For example, CNN is used to generate the deep features located in fully connected layers to predict the survival outcome \cite{lao2017deep}. While the CNN hidden layers with their feature-maps have been quantified and used for classification tasks as a panel of informative descriptors\cite{JBHI2, chaddad2019deep}. Despite the potential of CNN-based radiomics in classification, no study yet has explored the deep radiomic analysis in predicting the immune cell status and/or impact on survival outcome for patients with brain tumours (i.e., LGG and GBM). We propose in this study to extract the radiomic features from the activation maps of a pre-trained 3D CNN to solve this problem. This model extends the work of \cite{JBHI2}, where the CNN activations applied to CT scans were modelled by Gaussian mixture model (GMM). We hypothesise that this approach, applied to four MRI sequences, can encoded important characteristics of tumor heterogeneity that are associated with immune cell markers.
	
	% =======================================================
	% # II. Impact of traps on large signal characteristics #
	% =======================================================
	\section{Related work}
	Radiomics has been shown to be successful in the computer-aided diagnosis of different cancer diseases \cite{chaddad2019radiomics}. Specifically in the context of personalised medicine, the deep CNN algorithms have indicated an incredible capacity to identify complex high-dimensional data associations and have been adequately utilized for disease diagnosis and treatment planning \cite{mazurowski2019deep}. These models additionally showed the potential in survival analysis \cite{jing2019deep}. Regardless of their successfully applications using the natural image, CNNs have had a more limited achievement in clinical imaging, mainly because of the limited data-sets that in CNN models lead to serious overfitting. 
	A simple solution is to quantify the pre-trained CNN models for prediction tasks using information theory \cite{chaddad2019deep}, standard radiomic functions \cite{chaddad2019radiomics} and GMM parameters \cite{JBHI2}. In \cite{chaddad2019deep}, the quantifier functions are unable to provide full information as suggested in \cite{JBHI2} that the CNN has been separately trained on MRI scans and used to generate GMM descriptors. Inspired from \cite{JBHI2}, we assume that the samples observed are generated from a mixture of Gaussian distributions whose parameters (average, covariance matrix and weights) are learned using the Expectation Maximization (EM) \cite{dempster1977maximum}. In particular, to predict the status of immune cells and survival groups of patients, we applied the GMM directly to CNN feature maps of brain tumour images.
	
	\section{Methodology}
\begin{figure*}[ht!] %!t
		\centering
		\includegraphics[width=.95\linewidth]{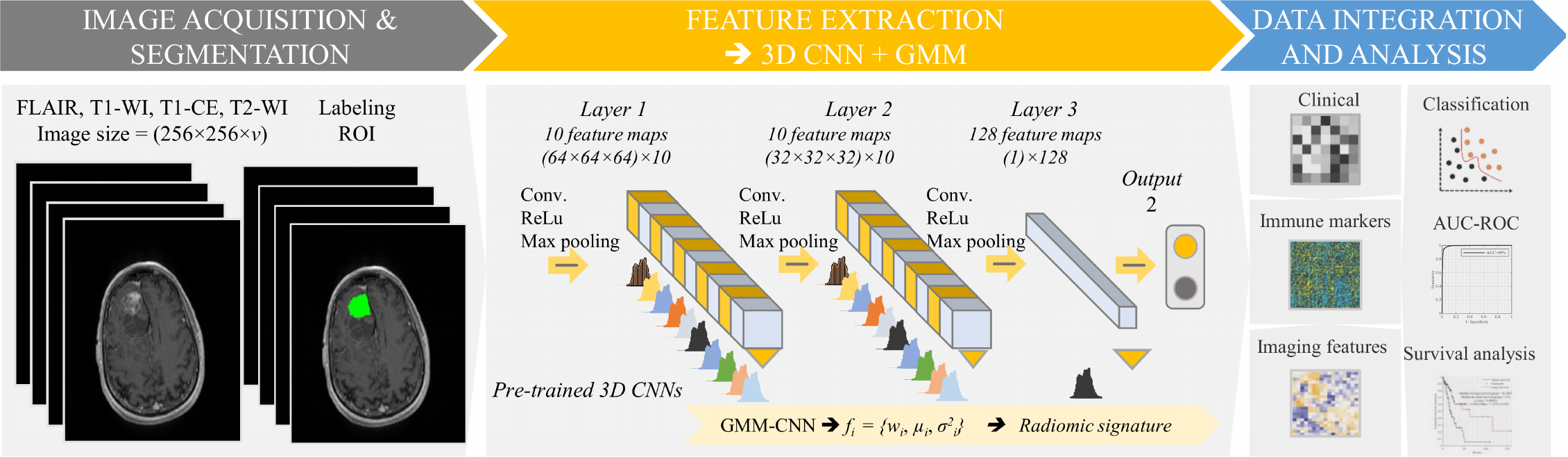}
		\caption{Deep radiomic workflow. 1) Four MRI sequences (T1-WI, T2-WI, T1-CE and FLAIR) are acquired for GBM and LGG patients; ROI is then delineated using the 3D Slicer tool. 2) A group of parametric features (i.e., average, variance and weight) quantify the GMM curves that fit the histogram of 3D CNN feature maps. 3) Aggregate the deep radiomic features with clinical variable and immune cell markers for analysis using the RF models, Kaplan-Meier estimator and log-rank test.}
		\label{FIG1}
	\end{figure*}

	Figure~\ref{FIG1} shows the pipeline of deep radiomic model. MRI is the first way to obtain T1-weighted (T1-WI), T1-weighted post-contrast/contrast enhancement (T1-CE), T2-weighted (T2-WI) and fluid-attenuated inversion recovery (FLAIR) imaging data from patients with brain tumor (i.e., LGG and GBM). In axial slice, the tumour region of interest (ROI) is semi-automatically delineated. For the generation of multi-scale texture from CNN features of each brain tumour, a pre-trained 3D CNN is then used. These CNN feature distributions is converted into elements vector of tumor heterogeneity using the GMM. Combined imaging features in a vector is corresponding  to immune  cell  markers  and clinical  information  (age,  gender  and  overall  survival). Last, we applied the random forests for predicting immune response status (low versus high, below or above median value of each of the three immune markers) and survival patient groups (short vs. long survival, below or above the median survival of patients). In addition, we performed the log-rank test and Kaplan-Meier (KM) estimator to evaluate the predicted survival groups from RF models. 
	
	\subsection{Patients and image pre-processing}
\begin{figure}[ht!] %!t
		\centering
		\includegraphics[width=.95\linewidth]{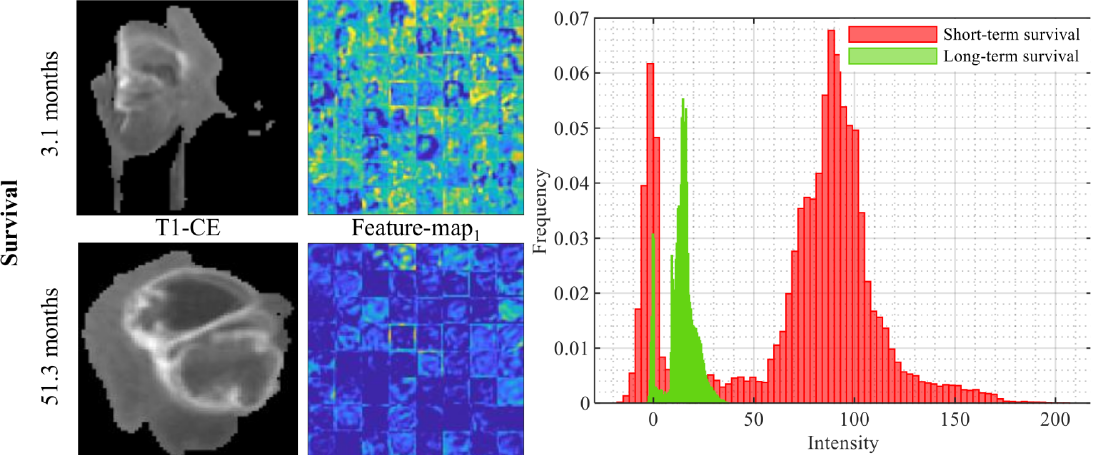}
		\caption{Example of short-term (top) and long-term (bottom) survival. (Left) ROI slices of post-contrast T1 MRI. (Middle) First feature-maps from the first convolutional layer and their histograms (Right).}
		\label{Fig2}
	\end{figure}

	Our study uses a dataset of 151 patients in the TCGA database with histologically confirmed with LGG (n=83) and GBM (n=68). Patients were chosen on the basis of the availability of high-quality T1-WI,  T2-WI, T1-CE and FLAIR scans and clinical variable (age , gender and overall survival) along with the corresponding immune cell markers. Images were obtained from The Cancer Imaging Archive (TCIA) \cite{prior2013tcia} that aggregates these 151 patients from multiple sites and thus the scanner model, pixel spacing, slice thickness and contrast varies within the selected cohort. We avoid these differences by sampling all images to a common voxel resolution of 1 mm$^{3}$ with a total size of 256 $\times$ 256 $\times$ slices voxels. Furhermore, intensities were standardised to the [0; 255] range. Immune cell markers (Macrophage M1, Neutrophils and T Cells Follicular Helper Immune cell markers) are the immune cellular fraction that was computed using CIBERSORT \cite{gentles2015prognostic}. More details on these immune markers are published in \cite{thorsson2018immune}. 
	The ROI in each MRI scan was labeled slice-by-slice using 3D Slicer\footnote{\url{https://www.slicer.org/}}. Using these ROIs, 3D CNN models is applied to generate CNN features that are estimated by their distributions using GMM.
	Figure~\ref{Fig2} shows two ROI slices for glioma with shorter-term (less than the median survival of 14.43 months) and longer-term (over 14.43 months) survival. In contrast to short-term ones (3.1 months), we see that long-term  survival (51.3 months) ROIs have more coarse-grained texture in the first activation map from the first convolutional layer of the network. We note that there is a distinct histogram of features from the same convolutional layer and feature maps between 3.1 months vs. 51.3 months and two Gaussian components may match their distributions. A similar conclusion was made when we evaluated the immune cell marker in which two distribution components could represent the histogram. 
	
	% =============================================
	% # III. Modeling and consistency validations #
	% =============================================
	\subsection{Deep feature computation}
	
	Deep CNN models could be used directly for classification tasks when large datasets are available. Unfortunately, CNN leads to overfitting when applied to small data sets. To overcome this issue, we propose to use pre-trained CNN and translate its feature-maps using GMM to a small set of distribution parameters (i.e., average, variance and weight of each mixture component) as previously mentioned.  The theory driving to use pretrained network on other datasets is that tumours such as LGG and GBM are primaly distinguished by local texture features. These features can then be effectively captured by agnostic CNN feature-maps. In this study, we used a 3D CNN that is pre-trained on 3D MRI for Alzheimer patients as described in \cite{chaddad2019deep}, with cross entropy loss,learning rate of 0.0005, and stochastic gradient descent optimization with momentum of 0.9. The details of the 3D CNN architecture are as follows.
	\textbf{Input}: image size = 64$\times$64$\times$64 voxels. \textbf{Layer 1}: filters=10; filter size = 2$\times$2$\times$2; stride=2; Max pooling; ReLU; dropout=0.8; output = 10 feature maps of size (32$\times$32$\times$32). \textbf{Layer 2}: filters=10; filter size = 2$\times$2 $\times$2; stride=2; Max pooling; ReLU; dropout=0.8; output = 10 feature maps of size (16$\times$16$\times$16). 
	\textbf{Layer 3}: fully connected layer; output = vector size 128. \textbf{Layer 4}: softmax = vector size 2. Using this model, the $i$-th feature map in layer $l$ is represented as a feature vector $\mathbf{f}$ with 3$\times k$ elements:
	\beq
	\mathbf{f}^{(l)}_{i} \ = \ \big[\mu_1, \sigma_1^2, \omega_1, \ldots, \mu_k,  \sigma_k^2, \omega_k\big].
	\eeq
	where each of the Gaussian components (i.e., $k$  is the number of components) is represented by average $\mu_j$, variance $\sigma^2_j$ and weight $\omega_j$.
	Considered $k=2$, with 21 3D feature-maps, the feature row vector $\mathbf{f}$ consist of 63$\times(k=2)$ features. The reason to consider the $k=2$ is representing by most histogram distribution of ROIs. However, other value of $k$ could be analysed. 
	
	\subsection{Classification and statistical analysis}
	The deep features are then used as input to RF model to predict the immune cell status (i.e., below or above median value of each immune cell marker) and classify the shorter-term from longer-term survival class (i.e., below or above median value of survival) of brain tumour. The hyper-parameters of RF models were chosen in our experiments using grid search on a validation set. 
	The imputation technique for adjusting the survival of the censored patients ( e.g. days to last visit or follow-up) was used to differentiate between survival groups (i.e. short and long survival) using the RF model, since the duration of the last visit only gives a lower limit on the true survival rank. In particular, each of the censored patients was given the average survival of uncensored subjects with a time-to-death greater than or equal to their own last follow-up time. 
	We considered the leave-one-out cross validation (LOOCV) for classification task, training images are split into $n$ samples, one sample is placed aside for testing at each time, and the remaining $n-1$ samples are used to train the RF classifier. 
	For the tested samples, the AUC value was then computed. Using Matlab's Statistics and Machine Learning Toolbox, all our processing and analysis steps were carried out. The Kaplan-Meier estimator and log-rank test are then used to assess the differences between predicted survival groups.

	\begin{table*}[ht!]
		\centering
		\caption{Summary of performance metrics for predicting the survival groups (Short-term and long term survival).}\label{table1}
		\setlength{\tabcolsep}{5pt}
		\renewcommand{\arraystretch}{1.1}
		\begin{small}
			\begin{tabular}{lcccccc}
				\toprule 
				\multirow[b]{2}{*}{\textbf{Features}} & 
				\multicolumn{2}{c}{\textbf{Median survival (month)}} & 
				\multirow[b]{2}{*}{\textbf{HR}} & 
				\multirow[b]{2}{*}{\textbf{CI}} & 
				\multirow[b]{2}{*}{\textbf{p-value}} & 
				\multirow[b]{2}{*}{\textbf{AUC}} \\
				\cmidrule(l){2-3} & \textbf{Shorter-term} & \textbf{Longer-term} \\ 
				\midrule
				Deep radiomic features (R) & 10.8  & 16.93 & 3.09 & 1.88\,--\,5.07 & 1.45 $\times$10$^{-5}$ & 73.01 \\
				Clinical variable (C)  & 13.9   & 14.73 & 1.59 & 0.98\,--\,2.59 & 0.07 & 62.51\\
				Immune cell markers (I)  & 13.95   & 14.76 & 2.10 & 1.29\,--\,3.40 & 0.003 & 65.54\\
				I+C  & 13.95   & 14.76 & 2.10 & 1.29\,--\,3.40 & 0.003 & 65.84\\
				R+C & 10.8  & 16.93 & 3.09 & 1.88\,--\,5.07 & 1.45 $\times$10$^{-5}$ & 73.51 \\
				R+I  & 10.68   & 17.20 & 3.83 & 2.30\,--\,6.37 & 4.31$\times$10$^{-7}$ & 73.91\\
				R+C+I  & 10.68   & 17.20 & 3.83 & 2.30\,--\,6.37 & 4.31$\times$10$^{-7}$ & 74.12\\
				\bottomrule
			\end{tabular}
		\end{small}\\
		AUC: Area under the ROC curve (AUC) derived from RF models,
		KM estimator and log-rank test analyses for the predicted survival groups of 151 brain tumor patients.
		HR: hazard ratio, CI: confidence intervals. +: refers to the combination.
	\end{table*}

	\section{Results}
\begin{figure}[ht!] %!t
		\centering
		\includegraphics[width=.95\linewidth]{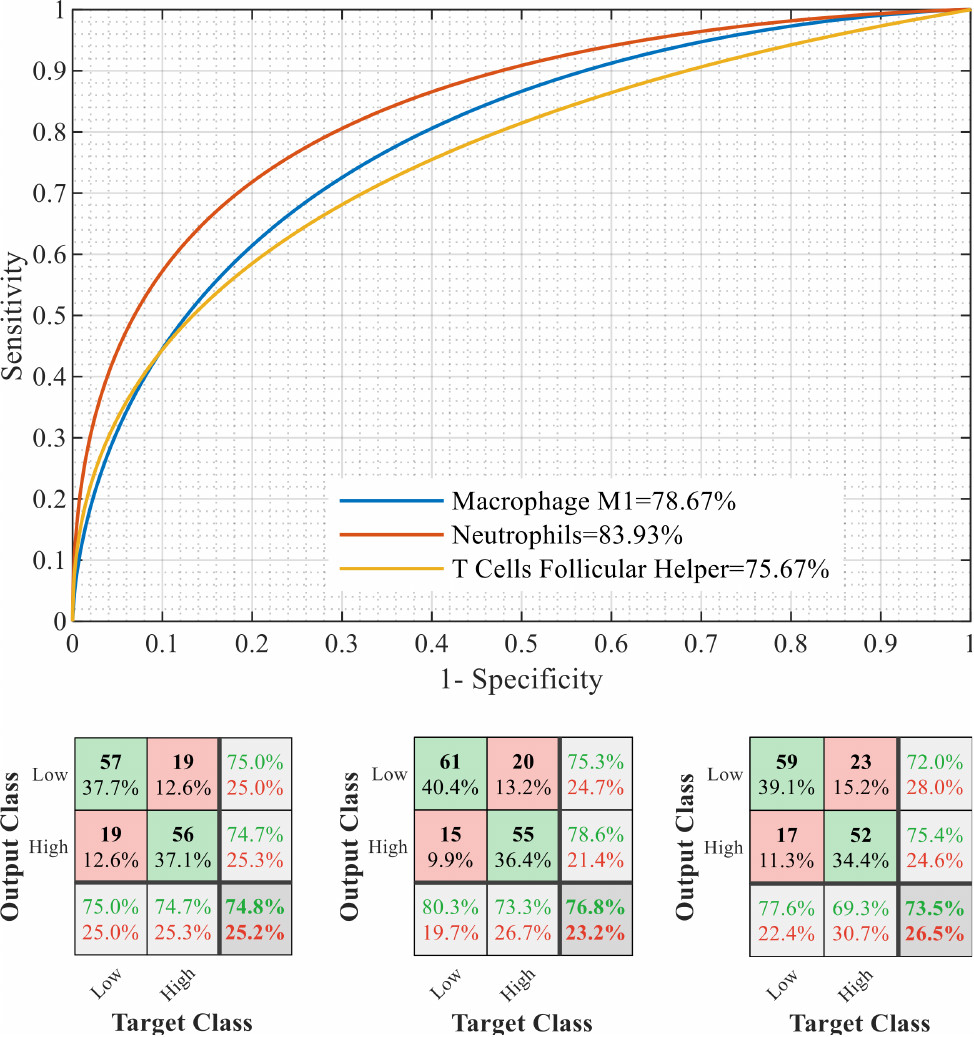}
		\caption{The results of AUC-ROC and confusion matrix (left: Macrophage M1, middle: Neutrophils, right: T Cells Follicular Helper) obtained from RF models use the deep radiomic features to predict the immune cell status of 151  brain tumors patients (LGG+GBM).}
		\label{FIG3}
	\end{figure}

	For classifying immune cell status, Figure \ref{FIG3} shows the AUC value (accuracy) of 78.67(74.8), 83.93(76.8) and 75.67(73.5)\% for Macrophage M1, Neutrophils and T Cells Follicular Helper, respectively. We found that the highest performance metrics (i.e., confusion matrix and AUC) were obtained to predict lower or higher values for the Neutrophil marker using deep features with RF models.
	
	\begin{figure}[ht!] %!t
		\centering
		\includegraphics[width=.95\linewidth]{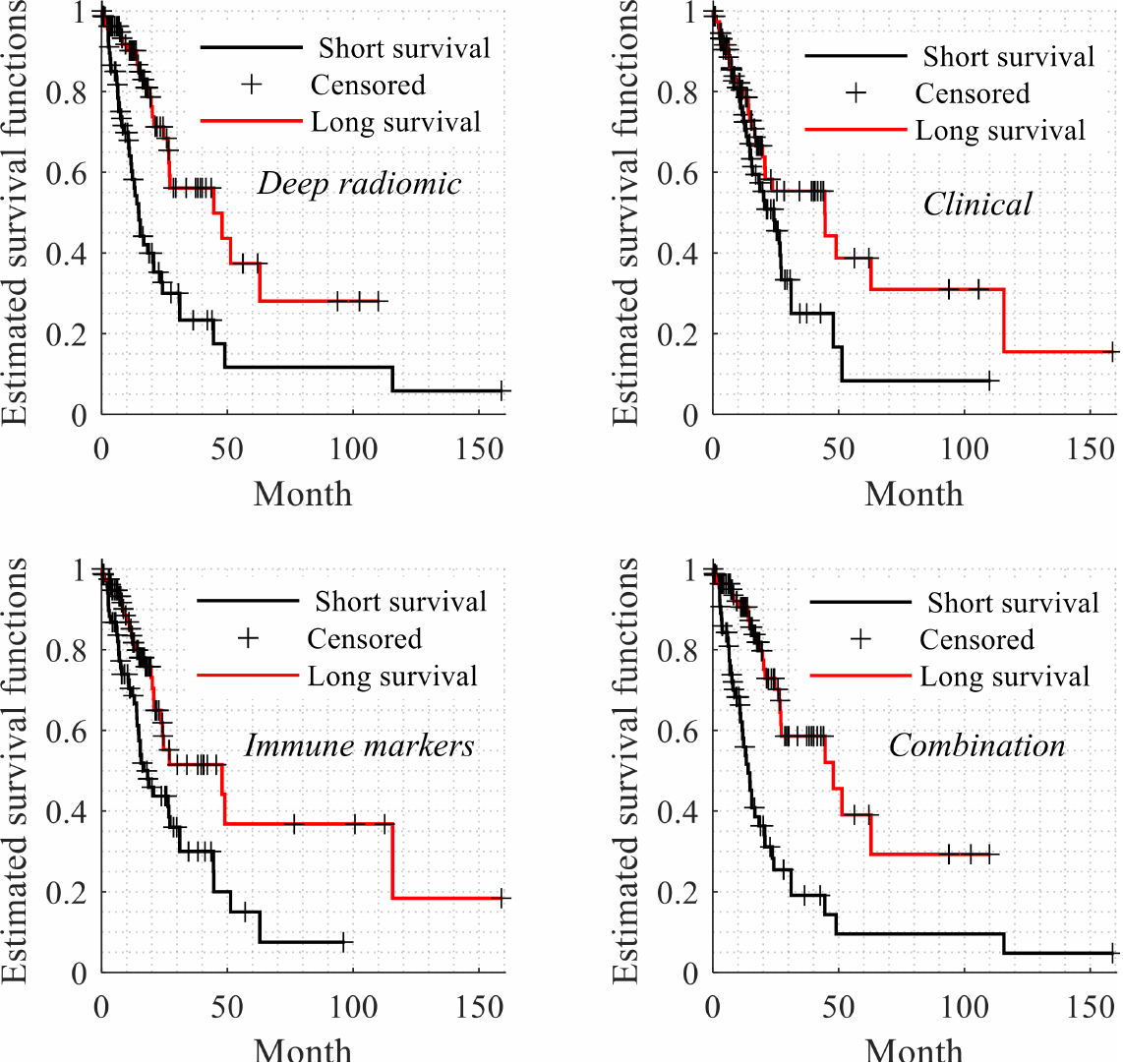}
		\caption{Log-rank test and KM estimator use deep radiomic features (R, $\mathbf{y}=$126 features), clinical variables (C: age and gender), immune cell markers (I: Macrophage M1, Neutrophils and T Cells Follicular Helper) and their combination (R+C+I) to compare survival between two predicted survival groups based on RF models.}
		\label{FIG4}
	\end{figure}
	
	Table \ref{table1} reports the performance metrics of survival analysis. For classifying short-term from long-term survival, RF model shows that the combination of deep radiomic features (R), immune cell markers (I) and clinical variables (C) is achieved a highest AUC of 74.12\% compared to 73.01\% for R (73.51\% for R+C), 65.54\% for I (73.91\% for R+I) and 62.51\% for C (65.84\% for I+C), respectively. 
	The predicted survival groups by RF models are then used to measure the significance by applying the Log-rank test and KM estimator. Figure \ref{FIG4} shows KM survival curves, except for clinical features for which the curves of shorter-term and longer-term survival overlap on ~ 20 months, all features give clearly-separated curves after a period of ~ 5 months. These findings are consistent with AUC values. Specifically, the combination of R+I+C leads to best significance p of 4.31$\times$10$^{-7}$ compared to the RF models using separately the R (p = 1.45 $\times$10$^{-5}$), I (p=0.003) and C (p= 0.07), respectively. We found that the immune cell markers and deep radiomic features are statistically significant for predicting the survival of patients with brain tumor. Considering k$\boldsymbol{\geq}$ 2, the predicted survival groups are insignificant with p$\boldsymbol{>}$0.05.

	% ==================
	% # IV. CONCLUSION #
	% ==================
	
	\section{Conclusion}
	This research examined deep radiomic descriptors that modelled 3D CNN features  using Gaussian mixture model to predict the immune cell makers and the survival in brain tumour patients. Our results suggest interactions between the deep radiomic signature, the markers of immune cells and overall survival. Inspired by these findings, we plan to extend the suggested deep radiomic pipeline through different types of cancer. 
	
	\section{Compliance with ethical standards}
	This research study was conducted retrospectively using human subject data made available in open access by TCGA and TCIA. Ethical approval was not required as confirmed by the license attached with the open access data.
	
	\section{Acknowledgments}
	This work was supported by Foreign Young Talents Program (No. QN20200233001). The funding agency has no role in the conceptualization of the study, data collection and analysis, or	the decision to publish these results. 
	
	%Research supported by Foreign Young Talents Program. The funding agency has no role in the conceptualization of the study, data collection and analysis, or the decision to publish these results.
	% ==================
	% # ACKNOLEDGMENTS #
	% ==================
	
	%\section*{Acknowledgment}

	% ==============
	% # REFERENCES #
	% ==============
	
	\bibliographystyle{IEEEtran}
	\bibliography{IEEEabrv,biblio_traps_dynamics}
	
\end{document}